# EMPIRICAL STUDY ON SELECTION OF TEAM MEMBERS FOR SOFTWARE PROJECTS – DATA MINING APPROACH

**SANGITA GUPTA[1], SUMA. V.[2]**

[1]Jain University, Bangalore
[2]Dayanada Sagar Institute, Bangalore, India

**Abstract-** One of the essential requisites of any software industry is the development of customer satisfied products. However, accomplishing the aforesaid business objective depends upon the depth of quality of product that is engineered in the organization. Thus, generation of high quality depends upon process, which is in turn depends upon the people. Existing scenario in IT industries demands a requirement for deploying the right personnel for achieving desirable quality in the product through the existing process. The goal of this paper is to identify the criteria which will be used in industrial practice to select members of a software project team, and to look for relationships between these criteria and project success. Using semi-structured interviews and qualitative methods for data analysis and synthesis, a set of team building criteria was identified from project managers in industry. The findings show that the consistent use of the set of criteria correlated significantly with project success, and the criteria related to human factors present strong correlations with software quality and thereby project success. This knowledge enables decision making for project managers in allocation of right personnel to realize desired level

## I. INTRODUCTION

The main objective of any software company is to provide quality software to its customers. The best of software system is bound to fail without the right people working on it. One of the ways to achieve highest level of quality in software system is through discovering knowledge for deployment of project personnel to a project by predicting their performance. The knowledge is hidden among the data set and it is extractable through data mining techniques. Present paper is designed to justify the capabilities of data mining techniques in context of software success by offering a data mining model for software companies to select the right personnel for their project. In this research, the classification method is used to evaluate project member's performance. By this task we extract knowledge that describes project member's' performance in the current project. It helps earlier identification of parameters related to human component resulting in better software quality and thereby project success. Software Engineering is a discipline that aims at producing high quality software through systematic, well-disciplined approach of software development. It involves methods, tools, best practices and standards to achieve its objective [5]. However software engineering is not only about tools and methods but also human aspect involved to work on it. Even the best of software system cannot be developed without correct team members. Therefore Human Aspect in Software engineering which is an important basis for software quality needs more understanding and deeper investigation. To achieve high quality software, it is essential to extract knowledge from the large dataset related to project members. The main purpose of knowledge infrastructure for project management is to provide information from past experience of the organization to improve the execution of new projects. To achieve this objective, the knowledge infrastructure has to compile and organize empirical data which is present in the systems and is available for use by project managers [6]. Consequently, the key elements of building knowledge infrastructure are collecting and organizing the knowledge, making it available through models, and reusing it to improve the execution of projects. In software development the main components can be broadly classified as human aspect-people and processes. Though processes have been well organized and developed, human aspect is still at preliminary stages for study. Without deep consideration into human aspect of software engineering even the best of processes will not give the desired quality. To accomplish the above-said objective of software quality, organizations are now looking deeply into human aspects using various techniques-some of them non parametric and some parametric data mining methods. However data mining techniques in software engineering have proved to be important tools for decision making of management. The data collected require proper method of extracting knowledge from large repositories for better decision making. Knowledge discovery in databases (KDD), often called data mining, aims at the discovery of useful information from large collections of data. The main functions of data mining are applying various methods and algorithms in order to discover and extract patterns of stored data [2]. Data mining and knowledge discovery applications have got a rich focus due to its significance in decision making and it has become an essential component in various organizations. Data mining techniques have been introduced into new fields of Statistics, Databases, Machine Learning and Pattern Recognition. There are increasing research interests in using data mining in every aspect of technology. Data Mining, concerns with developing methods that discover knowledge from data originating from empirical environments [2]. Data Mining uses many techniques such as Decision Trees, Neural Networks, Naïve Bayes, K-Nearest neighbour, and many others. Using these techniques many kinds of knowledge can be discovered such as association rules, classifications and clustering. The discovered knowledge can be used for prediction in diverse applications [2]. Section II has more references on applications.

The main objective of this paper is to use data mining methodologies to predict project members 'performance for the particular project. Data mining provides many tasks that could be used to study the project member's performance. In this research, the classification task is used to evaluate project member_s performance .There are many approaches that are used for data classification, the decision tree method is used here. Information like college percentile, experience, domain knowledge assessment, communication skills, reasoning skills, time efficiency etc was collected from the project management system for prediction of performance for that project. Organization of the paper is as follows:
Section II specifies the related work in the domains of data mining. Section III provides research methodology followed during this investigation. Section IV presents research work and technique development details. Section V indicates the results obtained by classification technique for effective project management. Section VI summarizes and concludes the paper.

## II. BACKGROUND AND RELATEDWORK

Software is has become part of every aspect of human life. The increasing demand of software has led to the progress of continual research in the areas of quality assurance and effective project management[6]. Data mining and pattern recognition techniques have proven as one of the established techniques for effective project management. Data mining has been used for many aspects of software projects like defect management, test analysis, code optimization etc[4]. Authors in [8] have used data mining for Bug Reports Classification using Text Data Mining. Data mining has also been used by authors in [9] for other domains like educational databases. The authors in [7] have developed a data mining framework based on decision tree and association rules to generate useful rules for personnel selection and retention based on several attributes of employee for high technology industry. Data mining, also popularly known as Knowledge Discovery in Database, refers to extracting or mining" knowledge from large amounts of data. The sequences of steps identified in extracting knowledge from data are shown in Figure 1

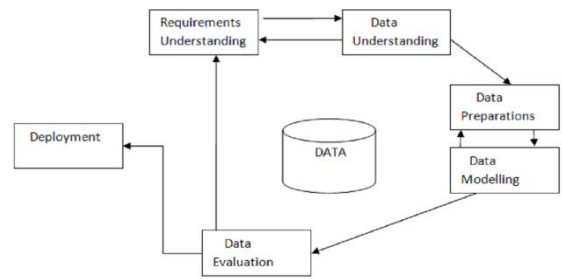

Figure 1-Research Methodology

Various algorithms and techniques like Classification, Clustering, Regression, Artificial Intelligence, Neural Networks, Association Rules, Decision Trees, Genetic Algorithm, Nearest Neighbour method etc., are used for data mining process [2]. Our Techniques and methods in data mining need brief mention to have better understanding.

Classification is the most commonly applied data mining technique, which employs a set of pre-classified examples to develop a model that can classify the population of records at large. This approach frequently employs decision tree or neural network-based classification algorithms. The classifier-training algorithm uses these pre-classified examples to determine the set of parameters required for proper discrimination. The algorithm then encodes these parameters into a model is called a classifier. The authors will be using decision tree for their research work.

A decision tree is a tree in which each branch node represents a choice between a number of alternatives, and each leaf node represents a decision. Decision tree starts with a root node on which it is for users to take actions. From this node, users split each node recursively according to decision tree learning algorithm. The final result is a decision tree in which each branch represents a possible scenario of decision and its outcome. Decision tree is tree-shaped structures that represent sets of decisions. These decisions generate rules for the classification of a dataset. Specific decision tree methods include Classification and Regression Trees (CART) and Chi Square Automatic Interaction Detection (CHAID).The authors in [1] have done a comparative study of the methods. The authors in [11] have developed many decision tree algorithm like ID3 and C5.The authors in [3] have investigated an incremental method for finding next node of the decision tree. The Decision Trees algorithm is a classification algorithm used for predictive modeling of multivariate attributes. For discrete attributes, the algorithm makes predictions based on the relationships between input columns in a dataset. It uses the values, known as states, of those columns to predict the states of a column that you designate as predictable. Specifically, the algorithm identifies the input columns that are correlated with the predictable column. The authors in [10] have used a Knowledge based Decision Trees algorithm which uses feature selection to guide the selection of the most useful attributes. In this study we have developed an algorithm to find the attributes using incremental method according to their mapping with performance. Thereafter a decision tree was constructed based on derived knowledge.

Data selection and transformation -In this step only those fields were selected which were required for data mining. A few derived variables were selected. While some of the information for the variables was extracted from the database. All the predictor and response variables which were derived from the database are given in Table I for reference. The huge data collected was thereby sampled and analyzed. Therefore, this work directed towards formulation of hypothesis for selection criteria of project personnel which has further impact on software quality of software projects. Modes of data collection include interactions with project developing team and human resource management. Empirical data analysis includes application of decision tree techniques to predict the efficiency of each project member for further deployment. The observational results indicate that though most software companies lay a lot of emphasis on general percentile aggregate but other factors like domain specific knowledge and reasoning skills play significant role for best performance in the organization. The most important factor which was analyzed was programming skills. Hereby the selection criteria have to be reframed giving more weightage to aspects like programming skills, depth in domain knowledge and reasoning skills rather than aggregate percentile.

Data mining and pattern recognition is gaining popularity because of its potentials to enhance our understanding and identifying, extracting and evaluating variables related to any process . By means of this method of classification method on multivariate attributes, it was found that the factors like project members 'programming skills, reasoning skills, communication skills; domain specific knowledge assessment and other attributes were highly correlated with the project member's performance rather than GPA.

### IV. RESEARCH WORK

PROPOSED TECHNIQUE In our research work, a data mining technique is used which is based on new Attribute Selection Measure function (heuristic) on existing C4.5 algorithm [10]. The drawback of C4.5 heuristic function (Gain ratio) is that, if the split information approaches zero, the ratio becomes unstable. In proposed technique for split criteria we have considered the maximum occurrences of each attribute value then calculating the average maximum occurrences of combination of each category attribute thus split information never reaches zero and gives more importance to realistic attributes and accurate results. The algorithm is as follows.

Step1. Let D, the Data partition be a training set of class labelled tuples. Suppose the class label attribute has m distinct values defining n distinct classes, $C_i$ ( for i=1,2----m). Let $C_{i,D}$ be the set of tuples of class $C_i$ in D. Let $|D|$ and $|C_{i,D}|$ denotes the number of tuples in D and $C_{i,D}$ respectively. Suppose attribute A on partition D having distinct values $a_1, a_2----, a_v, a_s$

### III. RESEARCH METHODOLOGY

This research focuses upon the selection of project personnel using classification technique for effective project management and thereby resulting in good software quality. In order to achieve the aforementioned objective, a deep investigation is carried out upon similar non critical projects from software industries to get the parameters for selection of project personnel.

Data Preparations -The training data set used in this study was obtained from software companies of Bangalore. Initially size of the data is 40. In this step data stored in different tables was joined in a single table after joining process, errors were removed. observed from the training data.

Step2. Calculate Attribute Selection Measurement Function (ASMF) for that attribute. Steps for calculating this function is as follows:

2.1 No. occurrences of each attribute value.
2.2 Occurrences of each category attribute.
2.3 Calculate $a_i/C_{iD}$

Step3. Compute average maximum occurrence for each attribute which denotes then $ASMF = \sum a_i/C_{iD}$.

3.1 Maximum occurrence of combination of each category and repeat Step 2
3. 2Select an attribute for which the value of ASMF is Maximum

Step4. Then on the basis of sorted values of ASMF we will divide the given training set into subsets and move to another level of tree.

Step5. Then we will repeat the same steps on each subset iteratively and derive a decision tree

Further sections will show data pre-processing and results after application of the researched technique.

## V. EMPIRICAL DATA ANALYSIS USING DECISION TREE ALGORITHM

Data was collected from a software company in Bangalore. It was preprocessed and prepared for analysis. It was subjected to data mining technique and the related to the hypothesis.

Data preparation is shown in Table 1– project personnel's attributes selected for analysis and Table 2-data with values of the attributes mentioned in Table 1.

TABLE I. Software project personnel related variables.

| Variable | Description | Values clustered for analysis |
|---|---|---|
| GPA | General Percentile aggregate | {good > 7.5<br>average >6.5 & <7.5<br>poor >5.0 & <6.5<br>} |
| DKA | Domain knowledge assessment | Poor ,Average, Good |
| PS | programming skills | Poor, average, good |
| GP | General Proficiency | Poor ,Average, Good |
| CS | Communication skills | Poor, average,good |
| TE | Time efficient | Poor ,Average, Good |
| RS | Reasoning skills | Poor ,Average, Good |
| P | Performance | Poor(3) ,Average(2), Good(1) |

The domain values for some of the variables were defined for the present investigation as follows:
All attributes marks are normalized out of 10 GPA – Previous institution marks
DKA – Domain Knowledge Assessment. Performance in domain knowledge assessment of the company. PS- programming skills results obtained by taking internal assessment on the programming concepts .
CS– Communication skills results obtained by seminar presentation of employee. Seminar performance is evaluated into four classes: Poor – Presentation and communication skill is low, Average – Either presentation is fine or Communication skill is average, Good – Both presentation and Communication skill is good,
RS– Reasoning skills. Reasoning skills performance
GP  - General Proficiency performance. Overall performance from previous project.
TE – Time efficiency of employee.
P – Performance

TABLE II. DATA-TRAINING SET

| S.No. | GPA | PS | DKA | CS | TE | RS | P |
|---|---|---|---|---|---|---|---|
| 1 | Good | Good | Good | Good | Good | Good | 1 |
| 2 | Good | Good | Average | Good | Good | Good | 1 |
| 3 | Good | Good | Average | Average | Average | Good | 1 |
| 4 | Average | Good | Good | Average | Good | Good | 1 |
| 5 | Average | Good | Average | Good | Average | Good | 1 |
| 6 | Poor | Good | Average | Average | Average | Average | 1 |
| 7 | Average | Good | Good | Good | Average | good | 1 |
| 8 | Poor | Good | Good | Average | Good | Average | 1 |
| 9 | Average | Good | Good | Good | Good | Good | 1 |
| 10 | Poor | Good | Good | Good | Good | Average | 1 |
| 11 | Good | Poor | Average | Poor | Average | Average | 2 |
| 12 | Good | Average | Average | Average | Poor | Good | 2 |
| 13 | Good | Average | Average | Good | Average | Good | 2 |
| 14 | Average | Average | Average | Good | Average | Good | 2 |
| 15 | Average | Average | Average | Good | Average | Average | 2 |
| 16 | Good | Average | Good | Good | Good | Good | 2 |
| 17 | Average | Average | Average | Poor | Average | Average | 2 |
| 18 | Average | Good | Average | Average | Average | Good | 2 |
| 19 | Average | Good | Average | Poor | Average | Average | 2 |
| 20 | Poor | Average | Average | Poor | Average | Poor | 2 |
| 21 | Average | Average | Good | Good | Good | Average | 2 |
| 22 | Poor | Average | Average | Good | Average | Average | 2 |
| 23 | Poor | Average | Average | Average | Average | Poor | 2 |
| 24 | Poor | Good | Average | Average | Average | Poor | 2 |
| 25 | Poor | Average | Good | Good | Good | Average | 2 |
| 26 | Poor | Average | Good | Good | Poor | Average | 2 |
| 27 | Average | Poor | Poor | Poor | Poor | Poor | 3 |
| 28 | Average | Poor | Poor | Average | Poor | Poor | 3 |
| 29 | Average | Poor | Poor | Poor | Poor | Poor | 3 |
| 30 | Average | Poor | Poor | Average | Poor | Poor | 3 |
| 31 | Good | Poor | Average | Average | Average | Good | 3 |
| 32 | Average | Poor | Poor | Good | Poor | Average | 3 |
| 33 | Average | Poor | Average | Good | Average | Average | 3 |
| 34 | Average | Poor | Average | Poor | Average | Poor | 3 |
| 35 | Average | Average | Poor | Average | Poor | Average | 3 |
| 36 | Poor | Average | Poor | Average | Poor | Poor | 3 |
| 37 | Poor | Poor | Average | Average | Average | Poor | 3 |
| 38 | Poor | Poor | Poor | Good | Poor | Poor | 3 |
| 39 | Poor | Poor | Poor | Poor | Poor | poor | 3 |
| 40 | Poor | Poor | Poor | Poor | Poor | poor | 3 |

The result obtained by the algorithm can be put in a form of a table which can also be called a confusion matrix. A confusion matrix is a table that shows the results of the classification experiment. (ai/Cid) is calculated by dividing the number of occurrences of ai in Cid. Best attribute is such that good maps to 1, average maps to 2 and poor maps to 3 in performance. Therefore the ideal matrix should be as

| attribute | Performance | | | total | ASMF |
|---|---|---|---|---|---|
| | 1 | 2 | 3 | | |
| good | x | 0 | 0 | x | x/x=1 |
| average | 0 | y | 0 | y | y/y=1 |
| poor | 0 | 0 | z | z | z/z=1 |

Total of ASMF for ideal attribute =3

The ASMF = 3 ideally when all good will perform good, all average will perform average and all poor

Empirical Study on Selection of team members for software projects – Data mining Approachwill perform poor. The confusion matrix for GPA and PS is shown below. Table III denotes the ASMF SCORE of all attributes.

Performance

| gpa | 1 | 2 | 3 | total |
|---|---|---|---|---|
| good | 3 | 4 | 1 | 8 |
| average | 4 | 6 | 8 | 18 |
| poor | 3 | 6 | 5 | 14 |

Therefore the ASMF (GPA) = 3/8 + 6/18 + 5/15= 1.07
Similarly for PS

Performance

| PS | 1 | 2 | 3 | total |
|---|---|---|---|---|
| good | 10 | 3 | 0 | 13 |
| average | 0 | 12 | 2 | 14 |
| poor | 0 | 1 | 12 | 13 |

Computing for all Attributes we get the following:

TABLE III -ASMF SCORE FOR ALL ATTRIBUTES

| ATTRIBUTE | ASMF |
|---|---|
| GPA | 1.07 |
| PS | 2.55 |
| DKA | 2.20 |
| CS | 1.25 |
| TE | 2.09 |
| RS | 1.79 |

Based on this computation-Table III we can derive decision tree with PS which has the highest ASMF as root node and other attributes further down in their order. One classification rule can be generated for each path from each terminal node to root node. Pruning technique was executed by removing nodes with less than desired number of objects and after tree pruning process we have the following rules:

**RULE 1** If (PS="GOOD") and (GPA="GOOD" or "AVERAGE") and (RS= " GOOD" or "AVERAGE")and (DKA= "GOOD" or"AVERAGE" ) and (CS="GOOD" or AVERAGE) then P = GOOD.

**RULE 2** If (PS= "AVERAGE") and (GPA="AVERAGE" or "GOOD") and (RS="GOOD" or "AVERAGE")and (DKA="GOOD") and (CS="GOOD"or"AVERAGE") then P= GOOD

**RULE 3** If (PS="GOOD") and (GPA="AVERGAGE" or "POOR") and (RS= "AVERAGE")and (DKA= "AVERAGE" ) and (CS="GOOD" or AVERAGE) then P = AVERAGE.

**RULE 4** If (PS= "AVERAGE") and (RS="GOOD" or "AVERAGE")and (DKA="AVERAGE") and (CS="GOOD"or"AVERAGE") then P= AVERAGE.

**RULE 5** If (PS = "POOR") then P=POOR

The implementation of the above study was done in TreePlan software also called DTREG and the results of classification by the author and the tool were matching. Table IV shows the results obtained from the software tool.

TABLE IV- IMPLEMENTATION RESULT

| ============ Input Data ============ |
|---|
| Input data file: C:\Program Files (x86)\DTREG\Examples\Iris.csv |
| Number of variables (data columns): 7 |
| Data subsetting: Use all data rows |
| Number of data rows: 40 |
| Total weight for all rows: 40 |
| Rows with missing target or weight values: 0 |
| Rows with missing predictor values: 0 |
| ===== Overall Importance of Variables ============ |

| Variable | Importance |
|---|---|
| PS | 100.000 |
| DKA | 44.232 |
| TE | 37.249 |
| RS | 26.780 |
| CS | 13.743 |
| GPA | 8.407 |

Finished the analysis at 29-Jun-2013 08:06:17
Analysis run time: 00:00.17

The data set of 40 employee used in this study which was obtained from software company in Bangalore was basis for our classification technique.. The result and rules obtained can classify project members into three classes of performance- good (should be deployed), average (can be deployed with training) and poor(should not be deployed).

Further scope of this research work is using other classification techniques and doing a comparative study. We can experiment on different set of attributes and find the most promising selection criteria.

## VI. CONCLUSION

In this paper, the classification task is used on project member's database to predict the project member's performance on the basis of previous database. As there are many approaches that are used for data classification, the decision tree method is used here. The resulting decision tree provides a representation of the concept those appeals to human because it renders the classification process self-evident. These parameters were collected from the employer's previous database, to predict the performance in the current project. It was noted that though the GPA may be low but other aspects like programming skills, domain knowledge and reasoning skills played an important role in selection of a candidate for the software project. High correlation was found between, programming skills effecting performance of personnel working in software companies. Errors occur by focus on whole but not in accordance to certain matrix. This study will help to the managers to improve the quality of software by deploying the right candidate. This study will also work to identify those human aspects which needed attention to reduce failure rate of software projects.

**ACKNOWLEDGMENT**

Authors would like to thank all the industry people who extended their valuable support and help in compliance within the framework of non-disclosure agreement.